\documentclass[pra,twocolumn,showpacs,superscriptaddress,aps]{revtex4}

\newcommand{\marker}[1]{\color{black}{#1}}

\usepackage{amsmath}
\usepackage{amssymb}

\usepackage{mathrsfs}
\usepackage{eucal}
\usepackage{mathbbol}
\usepackage{epsfig} 
\usepackage{graphicx}
\usepackage{color}
\usepackage{epstopdf}

\begin{document}

\title{Orientation dependence of energy absorption and relaxation dynamics of $\text{C}_{60}$ in fs-laser pulses: How round is $\text{C}_{60}$?}

\author{M. Fischer}
\affiliation{ Institut f\"{u}r Theoretische Physik, Technische Universit\"{a}t 
Dresden, D-01062 Dresden, Germany}
\affiliation{Max Planck Institute for the Physics of Complex Systems, 
N\"othnitzer Stra{\ss}e 38, D-01187 Dresden, Germany}
\author{J. Handt}
\affiliation{ Institut f\"{u}r Theoretische Physik, Technische Universit\"{a}t 
Dresden, D-01062 Dresden, Germany}
\affiliation{Theoretische Chemie, Technische Universit\"{a}t 
Dresden, D-01062 Dresden, Germany}
\author{G. Seifert}
\affiliation{Theoretische Chemie, Technische Universit\"{a}t 
Dresden, D-01062 Dresden, Germany}
\author{R. Schmidt}\email[]{E-mail: Ruediger.Schmidt@tu-dresden.de}
\affiliation{ Institut f\"{u}r Theoretische Physik, Technische Universit\"{a}t 
Dresden, D-01062 Dresden, Germany}

\date{\today}

\begin{abstract}
By means of non-adiabatic quantum molecular dynamics it is shown, that the amount of energy deposited into
$\text{C}_{60}$ by a short laser field strongly depends on the molecular orientation 
with respect to the laser polarization direction.
In consequence, subsequent electron-vibration coupling leads to different nuclear relaxation mechanisms with mainly three pathways:
(1)  excitation of giant $\text{A}_g(1)$ modes ("breathing"), (2) formation of deformed cage-like complexes ("isomers"),
fragmentation predominantly into two large pieces ("fission"). The results are in accord with and explain nicely already
existing experimental data. Future experiments are proposed to confirm the detailed predictions.
\end{abstract}


\maketitle

Since its discovery in 1985 \cite{Kroto1985}, the Buckminster fullerene $\text{C}_{60}$ has been investigated intensively 
in many fields of physics, chemistry and related areas.
With its well defined, highly symmetric structure and the large number of nuclear and electronic degrees of freedom (DOF),
$\text{C}_{60}$ has become an ideal model system to study structural, electronic and dynamical properties of atomic many-body systems,
such as electron transport in molecular systems \cite{Park2000}, collisions between complex particles \cite{Campbell2003}, cluster physics on surfaces \cite{Bonifazi2007}
or laser - finite matter interaction \cite{Laarmann2008}.
Due to its icosahedral symmetry (resembling a soccer ball~\cite{Kroto1985}),
one would intuitively expect a minor impact of orientation effects on the underlying mechanisms.
And indeed, many laser and collision induced phenomena have been explained succesfully within the spherical jellium approximation~\cite{Yabana1993,Thumm1995,Schlathoelter1999,Kidun2001,Verkhovtsev2012_1,Bauer2001,Scully2005,Madjet2008,Belyaev2009} or the infinitely conducting sphere model~\cite{Cederquist2000,Thumm1994}.

However, Gutierrez {\it et al.} \cite{Gutierrez2002} found, that the conductance across a $\text{C}_{60}$ junction 
between two carbon nanotubes changes with the orientation of the molecule with respect to the tubes over 
several orders of magnitude. Furthermore, it has been observed experimentally \cite{Glotov2001}, that the fusion cross section in
fullerene-fullerene collisions is several orders of magnitude smaller than that of the expected geometrical one. This is
due to the fact, that only very few and specific relative orientations between the colliding clusters contribute to fusion, as theoretically
confirmed by quantum molecular dynamics calculations \cite{Rohmund1996,Glotov2001}. 
A recent experiment by Daughton {\it et al.} \cite{Daughton2011} suggests, that the charge transfer to 
$\text{C}_{60}$ at surfaces depends on the molecular orientation as well. 
To what extent the molecular orientation plays a role in the laser-induced fullerene dynamics is, 
to the best of our knowledge, still an open question.

The response of $\text{C}_{60}$ to short intense laser pulses is subject of great current experimental~\cite{Tchaplyguine2000,Hertel2005,Laarmann2007,Shchatsinin2008,Bhardwaj04} and theoretical~\cite{Torralva2001, Frauenheim2002, Zhang2004,Zhang2006,Sahnoun06,Horvath2008,Beu2009,Tang2010,Bertsch1991} interest.
The most striking feature of fs-pulses (in contrast to ps-pulses) concerns the lack of small fragments and the predominant abundance of intact,
multiply charged fullerene ions in the experimental mass spectra of the photofragments \cite{Shchatsinin2008}. It has been argued \cite{Shchatsinin2008}, that most of the absorbed energy remains
in the electronic system preventing fragmentation even at intensities as large as 
$I \approx 10^{14} \frac{\text{W}}{\text{cm}^2}$.
This argument has been supported by pump-probe-control experiments where the excitation of the giant $\text{A}_g(1)$ "breathing" mode 
has been identified as one of the nuclear relaxation pathways \cite{Laarmann2007}. This vibrational mode assimilates only a few $10 \text{\ eV}$ of the total amount of electronically absorbed energy
of several $100 \text{\ eV}$ \cite{Laarmann2007}. However, a comprehensive understanding of the whole mechanism is still far from being reached.

In this report we study the energy deposition process and subsequent nuclear relaxation dynamics of $\text{C}_{60}$
in fs-laser pulses with strong focus on the orientation of the  molecule with respect to the laser polarization direction.
The investigations are based on the Non-Adiabatic Quantum Molecular Dynamics (NA-QMD) method \cite{Saalmann1996, Kunert2003a, Uhlmann2006a, Uhlmann2005}
which couples self-consistently classical nuclear motion with the quantum dynamics of electrons in the framework of 
time-dependent density functional theory. This universal many-body approach was proven in the past to reveal 
a large variety of collision- and laser-induced non-adiabatic phenomena in molecular systems ranging from dimers \cite{Uhlmann2003, Uhlmann2005b, Fischer2011, Fischer2012}, 
organic molecules \cite{Kunert2005, Handt2006}, 
metallic clusters \cite{Saalmann1998} up to systems as large as fullerenes \cite{Kunert2001, Laarmann2007} (for an overview see e.g. {\it www.dymol.org}). 
The large scale computations became possible by expanding the
time-dependent Kohn-Sham functions into a set of (atomic) basis functions (see \cite{Kunert2003a,Uhlmann2005} for details) which reduces the (still demanding) numerical effort
considerably, as compared to grid-based, non-adiabatic ab-initio methods~\cite{Reinhard1999, octopus}. 

\begin{figure}[!h]
\centering
\includegraphics[width=0.47\textwidth]{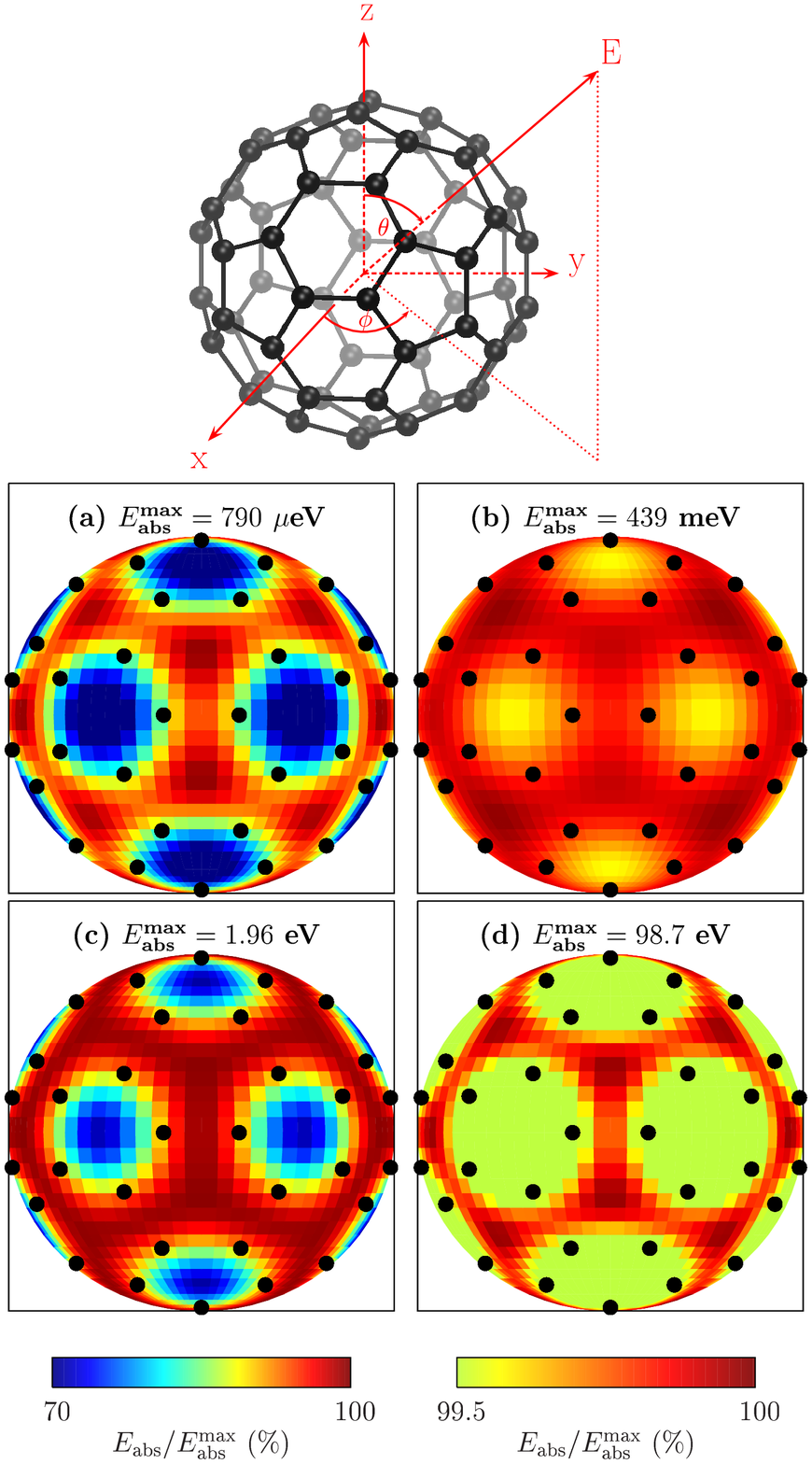}
\caption{(Color online) Orientation dependence of the absorbed energy $E_\text{abs}$ normalized to its maximum value $E_\text{abs}^\text{max}$ (given in each figure)
in a linearly polarized laser field {$\marker{\mathbf{E}(t;\theta,\phi)}$} with different laser parameters ($T = 10 \text{\ fs}$ in all cases): \\
 - left column (a),(c): $\lambda = 800\ \text{nm}$ (non-resonant) \\
 - right column (b),(d): $\lambda = 400\ \text{nm}$ (resonant) \\
 - first row (a),(b): $I = 1.0 \cdot 10^{13} \frac{\text{W}}{\text{cm}^2}$ \\
 - second row (c),(d): $I = 1.2 \cdot 10^{14} \frac{\text{W}}{\text{cm}^2}$ \\
The black dots indicate the positions of the carbon atoms {\marker{(view on yz-plane, cf. sketch on the top of the Figure, where we have slightly rotated
the molecule for a complete definition of the polarization respectively orientation angles)}}. The results have been calculated using the fixed-nuclei approximation.}
\label{fig:short}
\end{figure}

In this study, we use a basis constructed from the cc-pVDZ and DeMon Coulomb Fitting 
basis sets \cite{basis,Feller1996,Schuchardt2007} for the description of the carbon atoms and the adiabatic local density approximation (ALDA) \cite{Zangwill1980, Zangwill1981}
for the exchange correlation potential.
With this basis electrons are allowed to be excited into ionizing continuum states;
 the explicit inclusion of absorbing boundary conditions~\cite{Uhlmann2006a} have been omitted in the present study.
The two innermost electrons of each carbon atom are treated within the frozen core approximation.
The electric field of the laser is parameterized according to 
\begin{align}
 {\marker{\mathbf{E}(t;\theta,\phi)}} = E_0 &\sin^2{ \left( \frac{\pi t}{T} \right) } \cos{\omega t} \ \times \notag\\
&\left[\cos{\phi}\sin{\theta}\ \mathbf{e}_x + \sin{\phi}\sin{\theta}\ \mathbf{e}_y + \cos{\theta}\ \mathbf{e}_z \right] \notag
\label{eq:efield}
\end{align}
with amplitude $E_0$, total pulse length $T$, and fundamental frequency $\omega$.
The angles $\phi$ and $\theta$ describe the polarization direction of the electric field with respect to a fixed molecular position
or, equivalently, the different orientations of the molecule in a settled linearly polarized laser field {\marker{(see sketch on the top of Fig. \ref{fig:short})}}.

First, we focus on the orientation dependence of the absorbed energy $E_\text{abs}$ in very {\it short} laser pulses with total time duration of $T = 10 \ \text{fs}$.
In this case, the nuclear motion is practically frozen and the calculations can be and have been done keeping the nuclear positions fixed. On top of that, this  
fixed-nuclei approximation reduces the numerical effort drastically and, thus, allows us to study possible orientation effects systematically as a
function of the remaining laser parameters. 
We choose those of experimental relevance with wavelengths of $\lambda = 400 \ \text{nm}$ (resonant) and $\lambda = 800 \ \text{nm}$ (non-resonant)
\footnote{The first dipole-allowed transition of $\text{C}_{60}$ from the highest occupied $\text{HOMO}$ to the lowest unoccupied molecular orbital $\text{LUMO+1}$ can be excited by a laser pulse with wavelength $\lambda \approx 400 \ \text{nm}$. Obviously, a laser pulse with $\lambda = 800 \ \text{nm}$ leads to non-resonant excitation.
{\marker{Note, that the HOMO-LUMO transition is dipole-forbidden.}}}, 
and intensities, $I = \frac{E_0^2}{2}$, varied from weak ($I = 1.0 \cdot 10^{13} \frac{\text{W}}{\text{cm}^2}$) up to moderate ($I = 1.2 \cdot 10^{14} \frac{\text{W}}{\text{cm}^2}$).
The results of the calculations are summarized in Fig. \ref{fig:short}.

As can be seen from Fig. \ref{fig:short}, in all four cases, the energy absorption depends significantly on the orientation of the molecule with respect to the laser polarization
direction. Specific orientations (corresponding to the red areas in and between the hexagons in Fig. \ref{fig:short}) lead to maximum absorbed energy values with
$E_\text{abs}/E_\text{abs}^\text{max} \approx 100 \%$. Other orientations 
(corresponding to the blue, green, yellow areas within the pentagons in Fig. \ref{fig:short})
result in distinctly smaller values. Obviously, the effect is strongest for the non-resonant wavelength of $\lambda = 800 \ \text{nm}$ with a total variation of 
$E_\text{abs}/E_\text{abs}^\text{max}$ of about $30 \%$ (left column of Fig. \ref{fig:short}). 

To understand this behavior, we have performed complementary orientation dependent real-time linear response calculations \cite{Kunert2003a, Tsolakidis2002}. In this method, the molecule is exposed to a very short 
$\delta$-shaped laser pulse containing all frequencies $\omega$. The optical spectrum $S(\omega) \sim \text{Im}\{\alpha(\omega)\}$ 
(with $\alpha(\omega)$ being the linear polarizability) is obtained from the Fourier transform of the induced time-dependent electronic dipole moment \cite{Kunert2003a}.
Although these calculations suffer from poor statistics 
\footnote{Note, in order to obtain sufficiently $\omega$-resolved spectra, simulation times of the order of ps are necessary, compared to $10 \ \text{fs}$ 
simulation time to calculate the absorbed energy.}, the polarizability $\alpha(\omega)$ shows qualitatively a similar orientation dependence and frequency sensitivity
as shown for the absorbed energy in Fig. \ref{fig:short}. Consequently, one can clearly attribute the effect to the anisotropy and $\omega$-dependence of the linear polarizability
of $\text{C}_{60}$, at least for these short and relatively weak laser fields.

\begin{figure}[htb]
\centering        
	\includegraphics[width=0.47\textwidth]{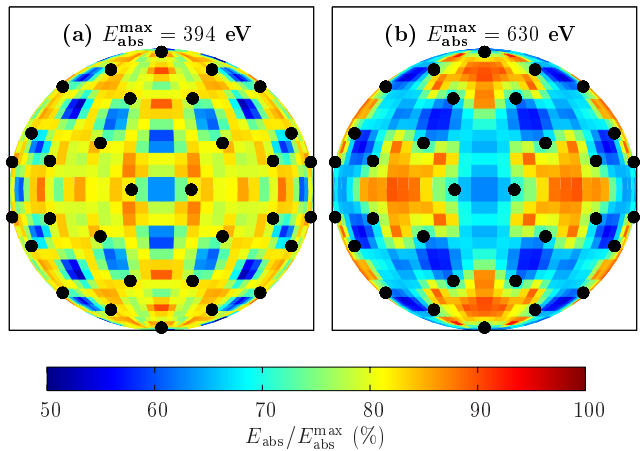}
\caption{(Color online) Same as in Fig. \ref{fig:short} but for a long ($T = 54 \text{\ fs}$), strong 
($I = 6.2 \cdot 10^{14} \frac{\text{W}}{\text{cm}^2}$) and non-resonant ($\lambda = 800 \text{\ nm}$) laser pulse:
(a) fixed-nuclei approximation, (b) full-dynamical calculations.}
\label{fig:long}
\end{figure}

\begin{figure}[htb] 
\centering
	\includegraphics[width=0.47\textwidth]{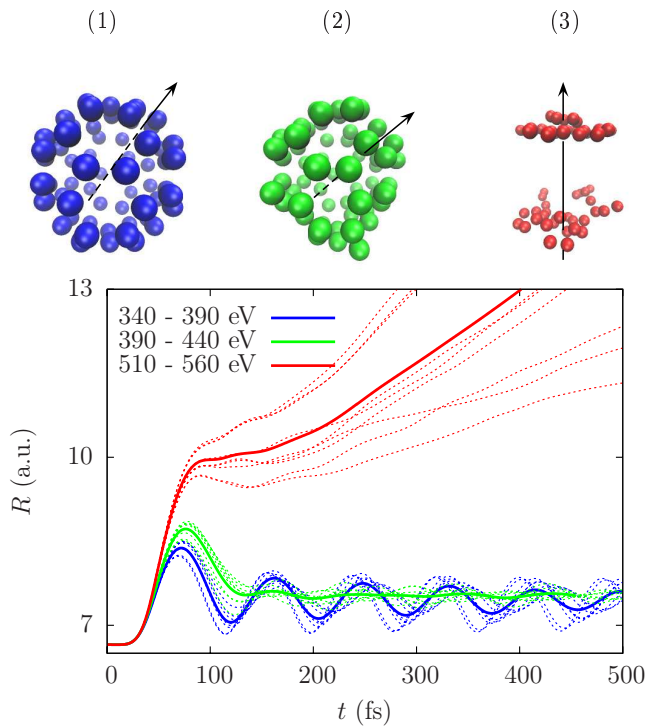}
\caption{(Color online) $\text{C}_{60}$ radius $R$ as a function of time $t$ for different molecular orientations (dotted lines) in the same laser field
($T = 54 \text{\ fs}$ , $I = 6.2 \cdot 10^{14} \frac{\text{W}}{\text{cm}^2}$, $\lambda = 800 \ \text{nm}$)
as used before (cf. Fig. \ref{fig:long}). The colors (blue, green, red) correspond to typical orientations within the same colored regions of Fig. \ref{fig:long} (b)
and, thus, belong to different energy absorption intervals given {\marker{in the Figure}}. The full lines are the corresponding mean values of $R(t)$ and
summarize the three different relaxation channels: (1) excitation of breathing modes (blue), (2) formation of non-breathing isomeric states (green), and (3) ultrafast almost symmetric fission (red).
{\marker{Typical snapshots of the dynamics at $t \approx 240 \text{\ fs}$ corresponding to the three relaxation channels are shown on the top of the Figure using the same colors. The black arrows
indicate the laser polarization axis [cf. Fig. \ref{fig:long} (b)].
Note, the fission process proceeds along the laser polarization direction.}}}
\label{fig:rad}
\end{figure}

Next, we consider in detail the orientation dependence of the absorbed energy $E_\text{abs}$ {\it and} the subsequent nuclear relaxation dynamics in a {\it long} 
($T = 54 \text{\ fs}$), {\it strong} ($I = 6.2 \cdot 10^{14} \frac{\text{W}}{\text{cm}^2}$) and non-resonant ($\lambda = 800 \text{\ nm}$) laser pulse. With a total pulse
duration of $T = 54 \text{\ fs}$, the nuclear motion will affect massively the absorption mechanism during the laser pulse (see below).
Above all, these laser parameters match also exactly those realized experimentally with highest intensity (Fig. 16 in ref. \cite{Hertel2005}).
The results of the calculations are summarized in Figs. \ref{fig:long} and\ref{fig:rad}.

In Fig. \ref{fig:long} (a), the absorption pattern of $E_\text{abs}$, again calculated within the fixed-nuclei approximation, is shown.
It is qualitatively different from that observed for the short pulses in  Fig. \ref{fig:short}, now with pronounced maxima within the pentagons and along the C-C single bonds between the pentagons
and hexagons. This pulse length dependence is maybe not so surprising, since it is well known already 
from a simple two-level system, that the population transfer between the two levels and, thus, energy absorption depend on the 
ratio between the pulse length and the intrinsic Rabi period. More importantly, the nuclear motion drastically influences the energy absorption
[cf. Fig. \ref{fig:long} (a) and (b)]. The maximum values of $E_\text{abs}$ are increased from $E_\text{abs}^\text{max} \approx 400 \text{\ eV }$ [Fig. \ref{fig:long} (a)]
up to about $E_\text{abs}^\text{max} \approx 600 \text{\ eV }$ [Fig. \ref{fig:long} (b)]. In addition, the orientation dependent profile of $E_\text{abs}$ is more
pronounced if nuclear motion is taken into account. As discussed below,
during the very initial stage, the radius of the cage is blowing up in all cases, which 
leads to a shift in the optical spectrum towards larger polarizabilities, resulting in a distinctly larger amount of absorbed energy.

In Fig. \ref{fig:rad} the average $\text{C}_{60}$
radius is shown on the sub-ps timescale for specific orientations corresponding to the blue, green and red colored regions in Fig. \ref{fig:long} (b) and, thus, to different intervals
of the absorbed energies {\marker{(given in Fig. \ref{fig:rad})}}. 
Clearly seen, during the laser pulse ($t \leq 54 \text{\ fs}$) the radius expands in all cases (corresponding to the first quarter of the 
"giant" $\text{A}_\text{g}(1)$ breathing mode). The initial excitation of the breathing mode 
is the key feature of the earliest stage of the electron-vibration coupling mechanism in $\text{C}_{60}$.

The long time behavior ($t \gtrsim 50 \text{\ fs}$) of the relaxation dynamics, however, depends on the orientation respectively on the amount of absorbed energy,
with three main channels {\marker{(see also Fig. \ref{fig:rad})}}:
\begin{enumerate}
 \item[(1)] {\it breathing:} For orientations with relatively low energy absorption (blue areas and blue dotted lines in Figs \ref{fig:long} and \ref{fig:rad}, respectively)
the breathing mode survives with amplitude and frequency depending on the actual electronic energy (cf. different dotted blue lines) in accord with our previous studies and
experiment \cite{Laarmann2007}. The present investigations deliver, in addition, a natural explanation for the weak pump-probe signal of this mode in the experimental
spectra: For a given laser pulse and $\text{C}_{60}$ in the gas phase, all orientations of the cage will contribute simultaneously and, thus, all nuclear relaxation channels
will do so (see below).
 \item[(2)] {\it isomers:} At intermediate absorbed energy (green) highly deformed but still stable complexes are formed with a nearly constant average $\text{C}_{60}$ radius $\sim 7.5 \ \text{a.u.}$
	(cf. the equilibrium $\text{C}_{60}$ radius $\sim 6.7 \ \text{a.u.}$). 
	These electronically excited isomeric states of $\text{C}_{60}$ are characterized by dominant excitations of vibrational ``surface'' modes which conserve the volume of the complexes.
	As in the former case (1), these isomers are also very stable {\marker{on the sub-ps timescale}},
        because only about $\frac{1}{10}$ of the total absorbed energy is transfered into vibrational energy (not shown).
        {\marker{However, on longer, i.e. ps (or ns), timescales one expects fragmentation, mainly via $\text{C}_2$ evaporation.}}
 \item[(3)] {\it fission:} For orientations with highest energy absorption (red) the cluster, however, fragments, basically into two, nearly equally sized large pieces (Fig. \ref{fig:rad}). 
            As seen from the behavior of $R(t)$ in Fig. \ref{fig:rad} the fission events are the direct consequence and continuation of the initially excited giant breathing mode.
            Evidently, the (nearly) symmetric fission channel is the energetically favored pathway of the nuclear relaxation. 
	{\marker{The strongly deformed cap-like fission fragments (Fig. \ref{fig:rad}) are highly unstable and are expected to decay into many small fragments $\text{C}_n$ ($n \lesssim 13$) on a ps timescale.}}
	{\marker{As a fingerprint of fission and the most important feature for its experimental verification, our calculations show that
this process evolves along the laser polarization direction (see Fig. \ref{fig:rad} and text below).}}
\end{enumerate}

These in fact somewhat unexpected findings explain, however, nicely the experimentally observed evolution of the photofragment mass distribution as function
of intensity (see discussion to Fig. 16 in \cite{Hertel2005}): Whereas for intensities below $I \lesssim 10^{14} \frac{\text{W}}{\text{cm}^2}$ only the "usual" dominance of 
$\text{C}_{60}$ ions is observed, at the highest intensity (as considered in our actual calculations), in addition, a large number of small fragments $\text{C}_n$
with $n \lesssim 13$ is found. Obviously, a decrease of the intensity would necessarily lead to the absence of the fission (and thus final multifragmentation) channels.

In summary, we have studied, for the first time, the orientation dependence of the excitation and relaxation mechanism of $\text{C}_{60}$ in linearly polarized laser fields.
In the framework of ab-initio NA-QMD calculations we found: \\
(i) The amount of deposited energy depends strongly on the molecular orientation with respect to the laser polarization axis. \\
(ii) The initial stage of the electron-vibration coupling is overwhelmingly dominated by a blow-up of the cage radius (corresponding basically to the excitation 
of the first quarter of the giant $\text{A}_\text{g}(1)$ breathing mode), independent on orientation. \\
(iii) The after-pulse nuclear relaxation mechanism depends dramatically on the molecular orientation exhibiting mainly three pathways:
(1) excitation of long-living breathing modes, (2) formation of stable, non-breathing isomers, and (3) fission into almost equal pieces.

The orientation dependence of the absorbed energy (i) can hardly be verified experimentally with $\text{C}_{60}$-targets in the gas phase. However, if $\text{C}_{60}$
is oriented on a surface (similar as in the experiments of Daughton {\it et. al.} \cite{Daughton2011}) we predict strong orientation dependent mass spectra of the ablated fragments as a function
of the laser polarization axis with respect to the surface.

The initial blow-up mechanism (ii) and the relaxation channels (iii) should be directly observable in future gas phase experiments.
{\marker{In particular, our calculations predict that the small fragments $\text{C}_{n}$ ($n \lesssim 13$) resulting from the fission channel will be preferentially emitted along the laser polarization axis. Thus, the fission channel can be identified by measuring the angular distribution of these fragments. Furthermore, the characteristic blow-up mechanism should be detectable in pump-probe experiments similar to that used to identify the breathing mode \cite{Laarmann2007}.}}

Altogether, the present investigations show that, albeit $\text{C}_{60}$ is highly symmetric with spherical nuclear geometry, it is by no means round.
The non-spherical electronic ``skin``, induced by the discreteness of the nuclear ``frame``, causes a rich variety of basically different, orientation dependent, laser-induced phenomena.

We gratefully acknowledge the allocation of computer resources from the ZIH 
of the Technische Universit\"at Dresden and the financial support by the Deutsche 
Forschungsgemeinschaft through the Normalverfahren as well as the
European Union and the Free State of Saxony through project No. 13857/2379-ECEMP.


\begin{thebibliography}{10}

\bibitem{Kroto1985}
H.~W. Kroto {\it et~al.}, Nature {\bf 318},  162  (1985).

\bibitem{Park2000}
H. Park {\it et~al.}, Nature {\bf 407},  57  (2000).

\bibitem{Campbell2003}
E. Campbell, {\em Fullerene Collision Reactions}, {\em Developments in
  Fullerene Science} (Springer, ADDRESS, 2003).

\bibitem{Bonifazi2007}
O.~E. Davide~Bonifazi and F. Diederich, Chemical Society Reviews {\bf 36},  390
   (2007).

\bibitem{Laarmann2008}
T. Laarmann, C.~P. Schulz, and I.~V. Hertel, {\em Progress in Ultrafast Intense
  Laser Science III}, Vol.~89 of {\em Springer Series in Chemical Physics}
  (Springer Berlin Heidelberg, ADDRESS, 2008), pp.\ 129--148.

\bibitem{Yabana1993}
K. Yabana and G.~F. Bertsch, Physica Scripta {\bf 48},  633  (1993).

\bibitem{Thumm1995}
U. Thumm, T. Bastug, and B. Fricke, Phys. Rev. A {\bf 52},  2955  (1995).

\bibitem{Schlathoelter1999}
T. Schlath\"olter, O. Hadjar, R. Hoekstra, and R. Morgenstern, Phys. Rev. Lett.
  {\bf 82},  73  (1999).

\bibitem{Kidun2001}
O. Kidun and J. Berakdar, Phys. Rev. Lett. {\bf 87},  263401  (2001).

\bibitem{Verkhovtsev2012_1}
A.~V. Verkhovtsev {\it et~al.}, Journal of Physics B: Atomic, Molecular and
  Optical Physics {\bf 45},  215101  (2012).

\bibitem{Bauer2001}
D. Bauer, F. Ceccherini, A. Macchi, and F. Cornolti, Phys. Rev. A {\bf 64},
  063203  (2001).

\bibitem{Scully2005}
S.~W.~J. Scully {\it et~al.}, Phys. Rev. Lett. {\bf 94},  065503  (2005).

\bibitem{Madjet2008}
M.~E. Madjet, H.~S. Chakraborty, J.~M. Rost, and S.~T. Manson, Journal of
  Physics B: Atomic, Molecular and Optical Physics {\bf 41},  105101  (2008).

\bibitem{Belyaev2009}
A.~K. Belyaev {\it et~al.}, Physica Scripta {\bf 80},  048121  (2009).

\bibitem{Cederquist2000}
H. Cederquist {\it et~al.}, Phys. Rev. A {\bf 61},  022712  (2000).

\bibitem{Thumm1994}
U. Thumm, Journal of Physics B: Atomic, Molecular and Optical Physics {\bf 27},
   3515  (1994).

\bibitem{Gutierrez2002}
R. Gutierrez {\it et~al.}, Phys. Rev. B {\bf 65},  113410  (2002).

\bibitem{Glotov2001}
A. Glotov, O. Knospe, R. Schmidt, and E. Campbell, The European Physical
  Journal D - Atomic, Molecular, Optical and Plasma Physics {\bf 16},  333
  (2001).

\bibitem{Rohmund1996}
F. Rohmund {\it et~al.}, Phys. Rev. Lett. {\bf 76},  3289  (1996).

\bibitem{Daughton2011}
D.~R. Daughton and J.~A. Gupta, Applied Physics Letters {\bf 98},  133303
  (2011).

\bibitem{Tchaplyguine2000}
M. Tchaplyguine {\it et~al.}, The Journal of Chemical Physics {\bf 112},  2781
  (2000).

\bibitem{Hertel2005}
I.~V. Hertel, T. Laarmann, and C.~P. Schulz, Advances In Atomic, Molecular, and
  Optical Physics {\bf 50},  219  (2005).

\bibitem{Laarmann2007}
T. Laarmann {\it et~al.}, Phys. Rev. Lett. {\bf 98},  058302  (2007).

\bibitem{Shchatsinin2008}
I. Shchatsinin {\it et~al.}, The Journal of Chemical Physics {\bf 129},  204308
   (2008).

\bibitem{Bhardwaj04}
V.~R. Bhardwaj, P.~B. Corkum, and D.~M. Rayner, Phys. Rev. Lett. {\bf 93},
  043001  (2004).

\bibitem{Torralva2001}
B. Torralva {\it et~al.}, Phys. Rev. B {\bf 64},  153105  (2001).

\bibitem{Frauenheim2002}
T. Frauenheim {\it et~al.}, Journal of Physics: Condensed Matter {\bf 14},
  3015  (2002).

\bibitem{Zhang2004}
G.~P. Zhang and T.~F. George, Phys. Rev. Lett. {\bf 93},  147401  (2004).

\bibitem{Zhang2006}
G.~P. Zhang and T.~F. George, Phys. Rev. B {\bf 73},  035422  (2006).

\bibitem{Sahnoun06}
R. Sahnoun {\it et~al.}, The Journal of Chemical Physics {\bf 125},  184306
  (2006).

\bibitem{Horvath2008}
L. Horv\'ath and T.~A. Beu, Phys. Rev. B {\bf 77},  075102  (2008).

\bibitem{Beu2009}
T.~A. Beu, L. Horv\'ath, and I. Ghisoiu,
  Phys. Rev. B {\bf 79},  054112  (2009).

\bibitem{Tang2010}
H. Tang, H. Li, Y. Dou, and W. Fang, Molecular Simulation {\bf 36},  986
  (2010).

\bibitem{Bertsch1991}
G.~F. Bertsch, A. Bulgac, D. Tom\'anek, and Y. Wang, Phys. Rev. Lett. {\bf 67},
   2690  (1991).

\bibitem{Saalmann1996}
U. Saalmann and R. Schmidt, Z. Phys. D {\bf 153--163},  15  (1996).

\bibitem{Kunert2003a}
T. Kunert and R. Schmidt, Eur. Phys. J. D {\bf 25},  15  (2003).

\bibitem{Uhlmann2006a}
M. Uhlmann, T. Kunert, and R. Schmidt, J. Phys. B {\bf 39},  2989  (2006).

\bibitem{Uhlmann2005}
M. Uhlmann, T. Kunert, and R. Schmidt, Phys. Rev. E {\bf 72},  036704  (2005).

\bibitem{Uhlmann2003}
M. Uhlmann, T. Kunert, F. Grossmann, and R. Schmidt, Phys.~Rev.~A {\bf 67},
  013413  (2003).

\bibitem{Uhlmann2005b}
M. Uhlmann, T. Kunert, and R. Schmidt, Phys.~Rev.~A {\bf 72},  045402  (2005).

\bibitem{Fischer2011}
M. Fischer {\it et~al.}, New Journal of Physics {\bf 13},  053019  (2011).

\bibitem{Fischer2012}
M. Fischer {\it et~al.}, Phys. Rev. A {\bf 86},  053821  (2012).

\bibitem{Kunert2005}
T. Kunert, F. Grossmann, and R. Schmidt, Phys.~Rev.~A {\bf 72},  023422
  (2005).

\bibitem{Handt2006}
J. Handt, T. Kunert, and R. Schmidt, Chem.~Phys.~Lett. {\bf 428},  220  (2006).

\bibitem{Saalmann1998}
U. Saalmann and R. Schmidt, Phys. Rev. Lett. {\bf 80},  3213  (1998).

\bibitem{Kunert2001}
T. Kunert and R. Schmidt, Phys. Rev. Lett. {\bf 86},  5258  (2001).

\bibitem{Reinhard1999}
P.-G. Reinhard and E. Suraud, Journal of Cluster Science {\bf 10},  239
  (1999).

\bibitem{octopus}
M.~A.~L. Marques, A. Castro, G.~F. Bertsch, and A. Rubio, Computer Physics
  Communications {\bf 151},  60  (2003).

\bibitem{basis}
{Extensible computational chemistry environment basis set database}.

\bibitem{Feller1996}
D. Feller, Journal of Computational Chemistry {\bf 17},  1571  (1996).

\bibitem{Schuchardt2007}
K.~L. Schuchardt {\it et~al.}, Journal of Chemical Information and Modeling
  {\bf 47},  1045  (2007), pMID: 17428029.

\bibitem{Zangwill1980}
A. Zangwill and P. Soven, Phys. Rev. Lett. {\bf 45},  204  (1980).

\bibitem{Zangwill1981}
A. Zangwill and P. Soven, Phys. Rev. B {\bf 24},  4121  (1981).

\bibitem{Tsolakidis2002}
A. Tsolakidis, D. S\'anchez-Portal, and R.~M. Martin, Phys. Rev. B {\bf 66},
  235416  (2002).

\end{thebibliography}
\end{document}